\documentclass[a4paper]{article}
\usepackage{xcolor}
\usepackage[utf8]{inputenc}
\usepackage[T1]{polski}
\usepackage{graphicx}\graphicspath{{img/}}
\usepackage{lmodern,fancyhdr,secdot,float,amsmath,amsfonts,amsthm,amssymb}
\usepackage{booktabs,tabularx}
\usepackage{subcaption}
\usepackage{graphicx}
\usepackage[margin=2cm]{geometry}
\usepackage{multicol}
\usepackage[hidelinks,unicode]{hyperref}
\usepackage{titlesec}
\titleformat{\section}{\centering\normalsize\bf}{\thesection.}{.5em}{\MakeUppercase}
\titleformat*{\subsection}{\bf\normalsize\selectfont}
\titleformat*{\subsubsection}{\bf\normalsize\selectfont}
\newcommand{\titlePL}[1]{\large\textbf{ #1}}
\newcommand{\titleEN}[1]{\normalsize #1}
\newcommand{\keywordsPL}[1]{\small\textbf{Słowa kluczowe:} #1}
\newcommand{\keywordsEN}[1]{\small\textbf{Keywords:} #1}
\newcommand{\abstractPL}[1]{\small\textbf{Streszczenie:} #1}
\newcommand{\abstractEN}[1]{\small\textbf{Abstract:} #1}

\definecolor{logo_color}{RGB}{40, 69, 166}

\begin{document}\thispagestyle{empty}\pagestyle{fancy}
\begin{minipage}[t]{0.5\textwidth}\vspace{0pt}%
\includegraphics[scale=0.9]{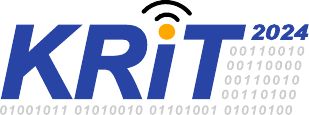}
\end{minipage}
\begin{minipage}[t]{0.45\textwidth}\vspace{12pt}%
\centering
\color{logo_color} KONFERENCJA RADIOKOMUNIKACJI\\ I TELEINFORMATYKI\\ KRiT 2024
\end{minipage}

\vspace{1cm}

\begin{center}
\titlePL{CHARAKTERYSTYKA ODBICIOWA INTELIGENTNYCH MATRYC ANTENOWYCH}

\titleEN{REFLECTION CHARACTERISTICS OF RECONFIGURABLE INTELLIGENT SURFACES}\medskip

 Paweł Hatka$^{1}$;
Dawid Brząkała$^{1}$;
Marcel Garczyk$^{1}$; 
Paweł Płaczkiewicz$^{1}$; 
Marta Sieradzka$^{2}$; 
Cyryl Prentki$^{1}$; 
Tomasz Jaworski$^{1}$; 
Krzysztof Cichoń$^{3}$; 
Adrian Kliks$^{3}$; 

\medskip

\begin{minipage}[t]{0.7\textwidth}
\small $^{1}$ Politechnika Poznańska, Poznań, \href{mailto:email}{imie.nazwisko@student.put.poznan.pl} \\
\small $^{2}$ Politechnika Poznańska, Poznań, \href{mailto:email}{marta.sieradzka.1@student.put.poznan.pl} \\
\small $^{3}$ Politechnika Poznańska, Poznań, \href{mailto:email}{imie.nazwisko@put.poznan.pl} \\

\end{minipage}

\medskip
\end{center}

\medskip

\begin{multicols}{2}
\noindent
\abstractPL{
W artykule przedstawiono wyniki pomiarów charakterystki odbiciowej dla rekonfigorowalnych matryc antenowych (RMA). Pomiary zostały przeprowadzone w środowisku nieidealnym, tj. typowym dla późniejszego praktycznego zastosowania. Podczas eksperymentów wykorzystano popularne matryce zrealizowane w projekcie otwartoźródłowym OpenSourceRIS. W badaniu skupiono się na uzyskaniu dwóch rodzajów charakterystki odbiciowej - dwu- i trójwymiarowej.}
\medskip

\noindent
\abstractEN{
This paper presents the results of reflection characteristics measurements for reconfigurable intelligent surfaces (RIS). The measurements were carried out in a non-ideal environment, i.e. typical for subsequent practical use of RISes. During the experiments, popular matrices implemented in the open-source project OpenSourceRIS were used. The study focused on obtaining two types of reflection characteristics - two- and three-dimensional.}
\medskip

\noindent
\keywordsPL{Rekonfigurowalne Matryce Antenowe, Charakterystyka Odbiciowa, Pomiary, 6G}
\medskip

\noindent
\keywordsEN{Reconfigurable Intelligent Surfaces, Reflection Characteristics, Measurements, 6G}

\section{Wprowadzenie}
Jednym z rozwiązań szeroko rozpatrywanych w kontekście rozwoju sieci bezprzewodowych jest zastosowanie rekonfigurowalnych matryc antenowych (RMA), znanych w literaturze anglojęzycznej jako Reconfigurable Intelligent Surfaces (RIS) \cite{Di2022,ETSI1}. Matryce te, zgodnie z założeniem, są kontrolowalnymi elementami umieszczonymi w środowisku propagacyjnym (tzn. pomiędzy nadajnikiem i odbiornikiem). Wspomniana możliwość zdalnej kontroli (np. przez operatora sieci bezprzewodowej) pozwala na zmianę charakterystki transmisyjnej środowiska radiowego w taki sposób, aby np. zapewnić uzyskanie oczekiwanych efektów propagacyjnych po stronie odbiornika \cite{Renzo2019}. W szczególności RMA pozwalać może na programowe definiowanie kąta odbicia padającego na matrycę sygnału (innego niż wynikałoby to z np. optyki geometrycznej) czy na częściową albo całkowitą absorpcję sygnału. Takie działanie może w konsekwencji skutkować poprawą określonych miar jakości sygnału po stronie odbiornika, np. zwiększenie mocy pożądanego sygnału odbieranego czy zminimalizowanie poziomu sygnałów interferencyjnych \cite{Liu2022}. \\
Macierze RMA są elementami określanymi często jako quasi-pasywne, tzn. same elementy antenowe są pasywne (w szczególności nie ma zastosowanego wzmacniacza), a zużycie mocy przez płytę RMA związane jest głównie z działaniem kontrolera \cite{ETSI1}. Macierze posiadają dużą liczbę elementów antenowych, co bezpośrednio przekłada się na możliwości oddziaływania RMA na sygnał padający. Każdy element można bowiem - w zależności od użytego rozwiazania - ustawić w różny tryb pracy. Im większa liczba składowych elementów antenowych tym większa możliwość uzyskania różnych wzorców odbiciowych dla takiej matrycy - jeżeli każdy element antenowy mógłby pracować w jednym z $M$ stanów (takim stanem jest np. wartość wprowawdzanego przesunięcia fazowego), to dla $N$-elementowej matrycy RMA można wygenerować $M^N$ różnych wzorców. Przestrzeń możliwości jest zatem bardzo duża. W praktycznym rozwiązaniu należałoby raczej skupić się na zdefiniowaniu bogatej, ale jednocześnie istotnie ograniczonej, książki kodowej wzorców, gdzie dla każdego wzorca przypisana byłaby odpowiednia charakterystyka odbiciowa płyty RMA. \\
Celem niniejszej pracy jest przedstawienie wyników pomiarów laboratoryjnych w środowisku typowym (tzn. nieidealnym) charakterystyk odbiciowych płyt RMA w dwóch wariantach - dwuwymiarowym oraz trójwymiarowym. Do celów pomiarowych wykorzystano matryce opracowane przez naukowców z Kolonii, zrealizowane w ramach projektu OpenSourceRIS \cite{Heinrichs2023}. 

\section{Środowisko Pomiarowe}
\label{rozdzial_2}
W pracy rozważono dwa sceniariusze pomiarowe. Celem pierwszego z nich było zmierzenie dwuwymiarowej charakterystyki odbiciowej RMA (charakterystyka horyzontalna). W drugim scenariuszu z kolei rozszerzono zakres pomiarowy o trzeci wymiar (wertykalny), uzyskując dzięki temu możliwość zbadania charakterystyki trójwymiarowej.
W celu przeprowadzenia wspomnianych badań w laboratorium zestawiono następujacy układ pomiarowy. Sygnał o częstotliwości 5.5 GHz oraz mocy -10~dBm wytwarzany był za pomocą generatora Rohde \& Schwarz SMM100A i transmitowany w kierunku macierzy RMA za pomocą anteny kierunkowej METIS 5.1-5.8~GHz. Po stronie odbiorczej zastosowano natomiast analizator widma Rohde \& Schwarz FSV3000, podłączony do takiej samej anteny. Poszczególne wartości nastaw analizatora są przedstawione w Tab. \ref{tab1}.
\begin{table}[H]
\centering
\caption{Nastawy analizatora widma}
\label{tab1}
\medskip
\begin{tabularx}{\columnwidth}{llX}
\toprule
\textit{Parametr} & \textit{Wartość} & \textit{Jednostka}\\ \midrule
Span & 0 & [Hz]\\
RBW & 500 & [Hz]\\
Sweep & 50 & [ms]\\
Reference level & -30 & [dBm]\\
Częst. środkowa & 5.5 & [GHz]\\
Poziom szumów & -110 & [dBm]\\ 
Tryb  & MaxHold & \\\bottomrule
\end{tabularx}
\end{table}

\begin{figure}[H]
\centering
\includegraphics[height=0.9\linewidth]{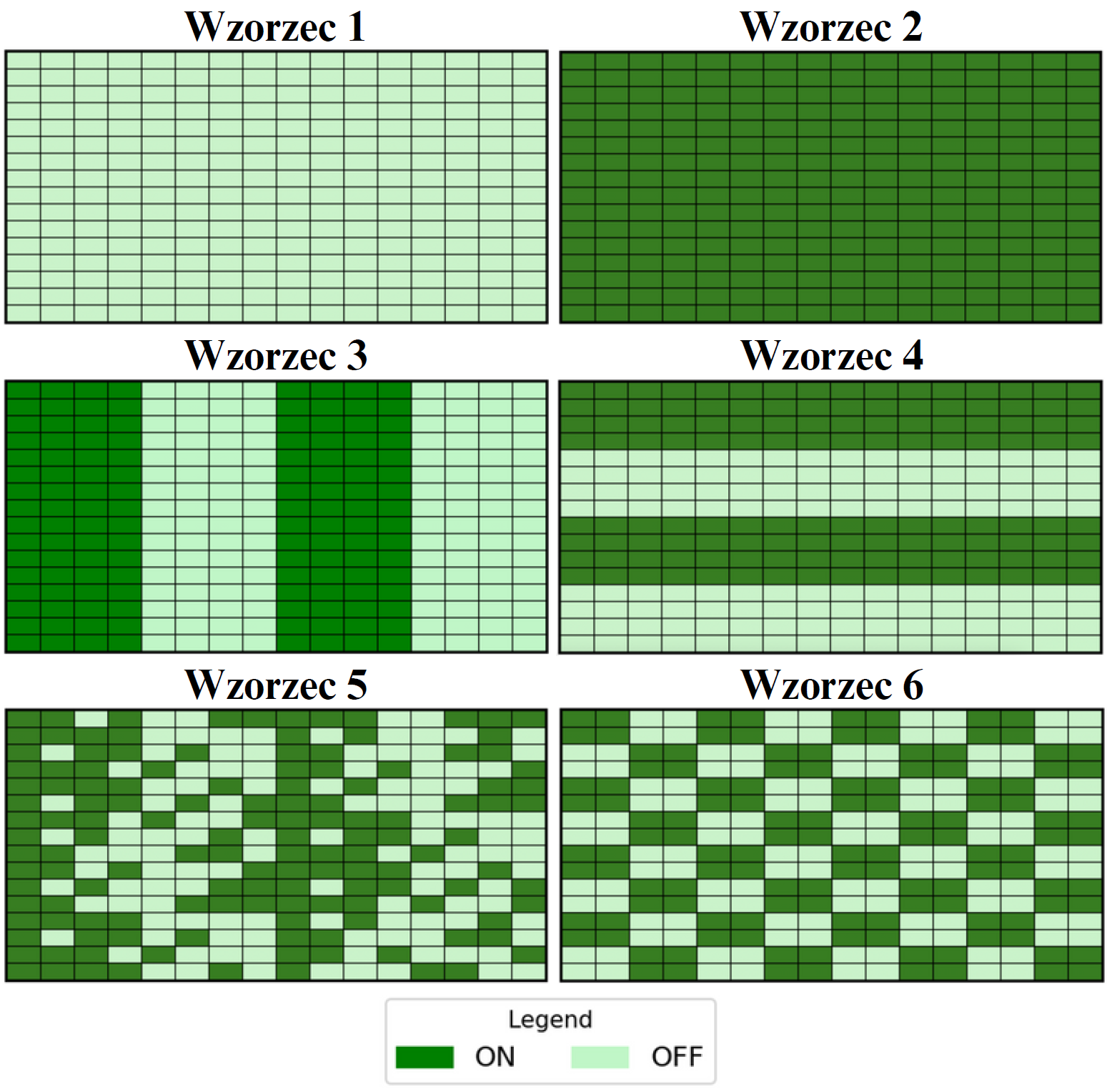}
\caption{Wizualizacja wzorców.}
\label{wizu_patternow_1_2_17_19_26_21}
\end{figure}

Obie anteny skierowane były na macierz RMA, dzięki czemu możliwe było zmierzenie poziomu mocu odbitej obserwowanej w odbiorniku. W celu oceny dwu- i trójwymiarowej charakterystyki odbiciowej należało zapewnić możliwość obrotu matrycy w płaszczyście poziomej i pionowej. Podczas ekpsperymenty zmiana kąta azymutu i elewacji umożliwiło specjalnie zaprojektowane i wykonane ramię wraz z silnikiem i układem sterującym.
Matryca RMA posiada $16 \times 16$ elementów odbijających, z których każdy jest indywidualnie sterowany za pomocą jednego bitu, tzn. każdy element może odbić padający sygnał bez zmiany jego fazy, ale może także zmienić ją o 180 stopni. Należy podkreślić, że nawet tak prosta zmienność stwarza bardzo szeroki wachlarz wariantów możliwych do wygenerowania, tj. $2^{256}$.
Cały proces pomiaru został zautomatyzowany przy pomocy oprogramowania napisanego w języku Python, które umożliwiło zdalne sterowanie analizatorem widma, generatorem sygnałowym, głowicą ramienia przytrzymującego matrycę, oraz zmianę wzorca ustawionego na RMA \cite{GitHub}. Wspomniany wzorzec określa, które z 256 elementów matrycy zmieniają fazę padającego sygnału o 180 stopni (dla f $ \approx $ 5.5GHz), tzn. które elementy są nieaktywne. Wspomniane 256 elementów zostało podzielone na 64 grupy każda po cztery elementy; w jednym wierszu są więc cztery podgrupy. Sterowanie matrycą polega więc na wybraniu, które z czterech elemntów w jednej z grup mają, a które nie mają odwracać fazy. Numeracja grup zaczyna się od indeksu 0, a pierwsza grupa znajduje się w lewym górnym rogu matrycy. 
Wybór realizowany jest poprzez wysłanie do RMA ciągu liczb w kodzie szesnastkowym, których pozycja w tym ciągu określa daną grupę. Zatem
0 - 0000 oznacza, że wszystkie elementy danego wiersza są wyłączone, F - 1111, że są włączone, a 3 - 0011, że dwie ostanie włączone itd. Przykładowy ciąg:
!0X800...000 - oznacza że zostanie włączony tylko pierwszy element grupy 0 \cite{GitHub1}. Przeprowadzono pomiary dla 27 wybranych wzorców, lecz ze względu na ograniczenia pojemności artykułu, w pracy zostaną przedstawione wyniki dla wybranych wzorców przedstawionych na Rys.~\ref{wizu_patternow_1_2_17_19_26_21}.

\subsection{Charakterystyka w jednej płaszczyźnie}
\begin{figure}[H]
\centering
\includegraphics[width=0.8\linewidth]{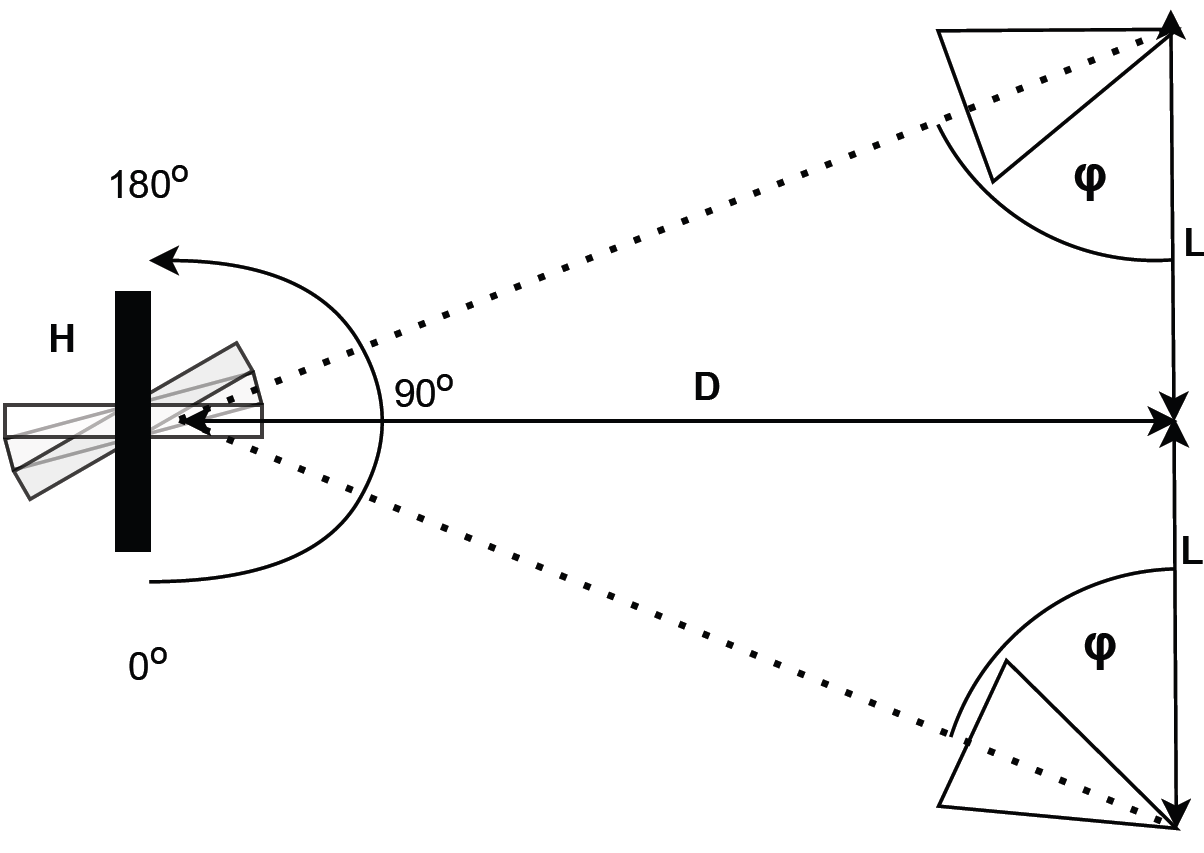}
\caption{Schemat pomiaru charakterstyki odbiciowej w jednej płaszczyźnie.}
\label{schemat_pomiaru_2D}
\end{figure}

Pomiary charakterystyki w płaszczyźnie poziomej w jakiej RMA odbija fale radiowe zostały wykonane zgodnie ze schematem (Rys.~\ref{schemat_pomiaru_2D}). 
Jak wspomniano, RMA została zamontowana na obrotowej głowicy umożliwającej ustawianie urządzenia pod zdanym kątem azymutu $\phi$ względem anteny nadawczej oraz odbiorczej. Odległość pomiędzy linią anten a linią matrycy RMA wynosi $D = 1.5$~m, natomiast odległość pomiędzy antenami wynosiła $2*L = 2$~m, co w wyniku daje kąt skierowania anten na RMA $\varphi = 56.3^\circ$. W celu otrzymania dokładnych wyników głowica wraz z zamontowaną matrycą wykonywała obrót z krokiem $1.8^\circ$ w płaszczyźnie azymutu, zaczynając od kąta $0^\circ$ a kończąc na kącie $180^\circ$. Osiągając kąt $90^\circ$ płaszczyzna matrycy była skierowana równolegle do lini anten. Anteny i matryca były umieszczone na wysokości 1.3~m.

\subsection{Charakterystyka trójwymiarowa}
\begin{figure}[H]
\centering
\includegraphics[width=0.8\linewidth]{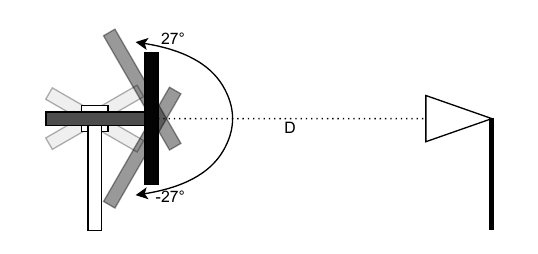}
\caption{Schemat pomiaru charakterystyki przestrzennej.}
\label{schemat_pomiaru_3D}
\end{figure}

Pomiary charakterystki przestrzennej zostały wykonane w sposób analogiczny dla charakterystyki 2D, jednak w tym przypadku wykorzystany został również obrót w płaszczyźnie elewacji zgodnie ze schematem przedstawionym na Rys.~\ref{schemat_pomiaru_3D}. Pomiary wykonane zostały z rozdzielczością 1.8{°} od 0{°} do 180{°} w płaszczyźnie azymutu oraz 9{°} od -27{°} do 27{°} w płaszczyźnie elewacji. Rozmieszczenie poszczególnych urządzeń wykorzystanych do przeprowadzenia pomiaru było identyczne jak w przypadku charakterystyki w jednej płaszczyźnie.

\section{Wyniki pomiarów}
\subsection{Scenariusz pierwszy}
\begin{figure}[H]
\centering
\includegraphics[width=1\linewidth]{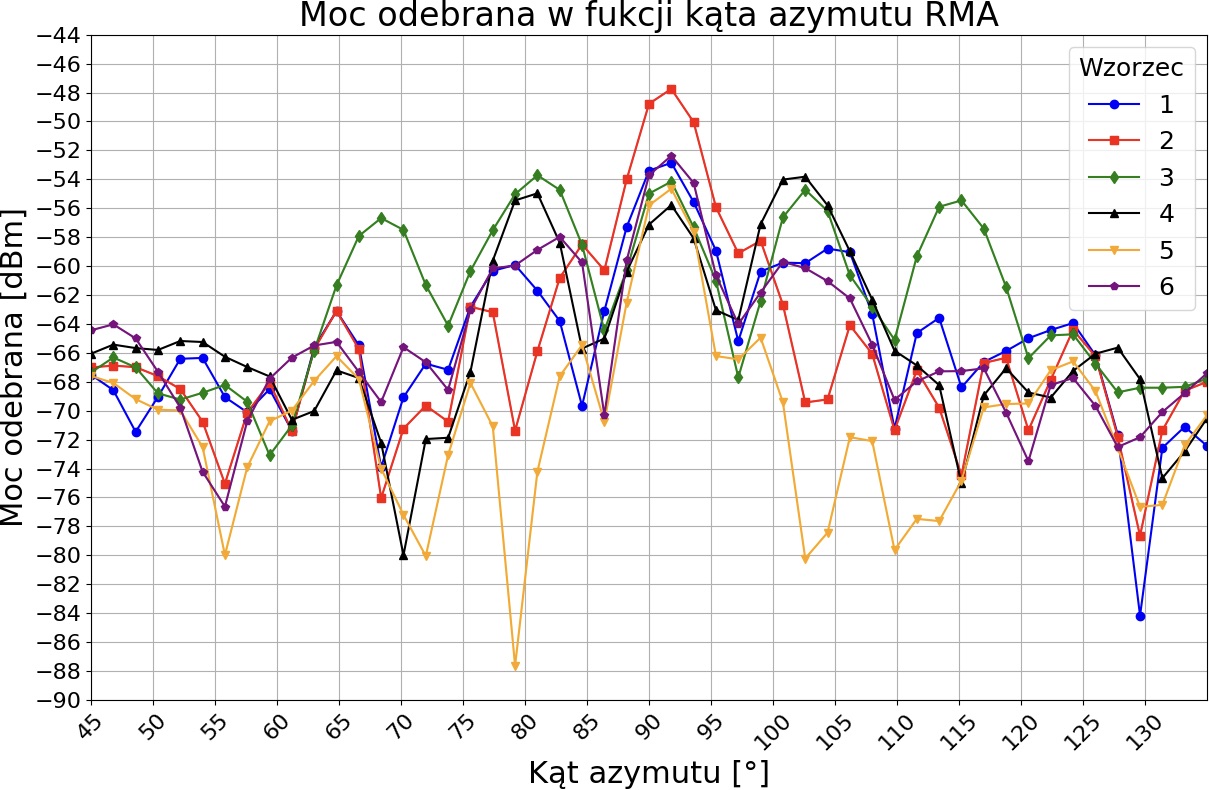}
\caption{Wyniki pomiarów charakterystki odbiciowej}
\label{2D_pattern}
\end{figure}

Na Rys.~\ref{2D_pattern}  przedstawiono wyniki pomiarów charakterystyki odbiciowej w jednej płaszczyźnie dla wzorców przedstawionych na Rys.~\ref{wizu_patternow_1_2_17_19_26_21}.
Dla wzorca pierwszego, zgodnie z oczekiwaniami, maksimum mocy odebranej występuje dla kąta 90°, zatem dla sytuacji, w której RMA wiązkę padającą odbija pod kątem równym kątowi padania. Wykorzystując wzorzec drugi widoczny jest wzrost mocy odebranej dla kąta 90° o około 5dB względem wzorca pierwszego, efekt ten wynika z aktywowania elemntów macierzy. Dla pozostałych kątów zauważalny jest brak poprawy albo wręcz spadek mocy odebranej. Dla  wzorca 3 i 4 maksimum charakterystyki występuje dla kąta innego niż 90°. Wzorzec trzeci przedstawia sytuację, gdzie elementy RMA były naprzemiennie włączone/wyłączone tworząc szyk grubych pasów pionowych. Dla takiego ustawienia RMA oraz pionowej polaryzacji anten zostały uzyskane interesujące wyniki w postaci kilku maksimów odbieranej mocy sygnału, oddzielonych "zerami", które pojawiały się co 10°-12° kąta azymutu. Podobna sytuacja występuje dla wzorca czwartego czyli grubych pasów poziomych, gdzie uzyskujemy trzy maksima dla kątów azymutu 80°, 90° oraz 105°.
W przypadku wzorca 5, możemy zaobserwować natomiast kilka punktów charakterystyki, dla których moc odebrana jest znacznie niższa, niż w przypadku innych wzorców. Możemy zaobserwować, że charakterystyka nie jest również symetryczna dla tego przypadku. Wyniki dla wzorca 6 ukazują występowanie przedziałów, w których charakterystyka wypłaszcza się, a poziom mocy odbieranej w całym zakresie kątowym różni się maksymalnie o ok. 2~dB. Pierwsze takie miejsce występuje miedzy kątem 59° a 66°, a drugie między 77.5° a 85°. Dla wszystkich wzorców poza 5 - losowym, kolejne ekstrema charakterystyki występują często mniej więcej symetrycznie względem położenia 90°, choć ich wartości co do mocy nie sią identyczne ze względu na charakter środowiska pomiarowego.

\subsection{Scenariusz drugi}
\begin{figure}[H]
\centering
\includegraphics[width=0.80\linewidth]{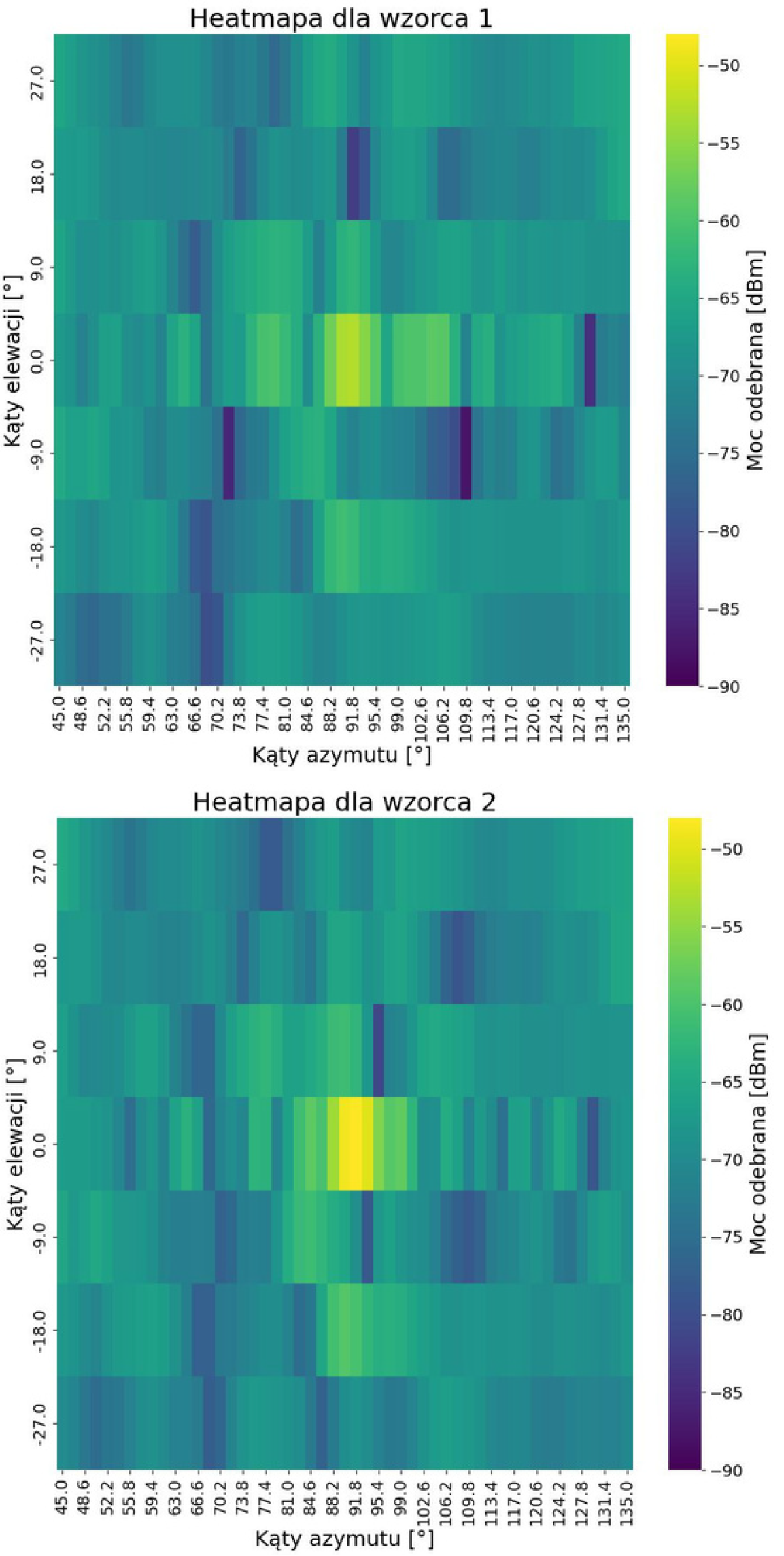}
\caption{Wyniki pomiarów charakterystki odbiciowej dla różnyh kątów elewacji dla wzorców 1 i 2.}
\label{wzorce12}
\end{figure}

\begin{figure}[H]
\centering
\includegraphics[width=0.79\linewidth]{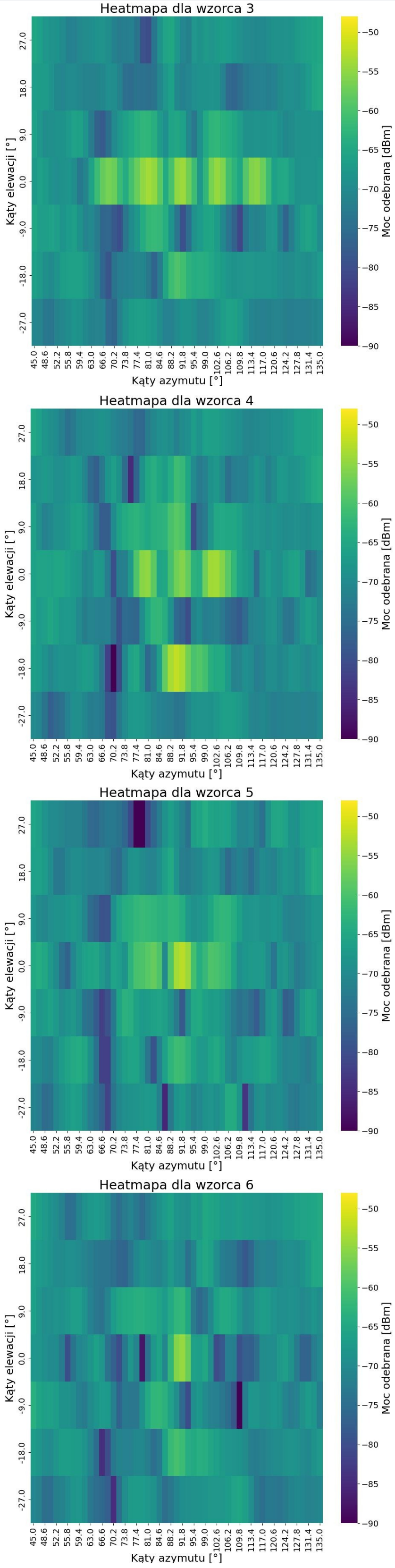}
\caption{Wyniki pomiarów charakterystki odbiciowej dla różnyh kątów elewacji dla wzorców 3 - 6.}
\label{wzorce3456}
\end{figure}

Otrzymane wyniki z pomiaru charakterystyki 3D zostały przedstawione w formie wykresów typu \textit{heatmap} (Rys.~\ref{wzorce12}-\ref{wzorce3456}). Analiza wyników scenariusza drugiego skupia się na kątach elewacji innych niż 0 stopni ze względu na wyniki pokrywające się ze scenariuszem pierwszym. Dla każdego z analizowanych przypadków można zaobserwować spadek mocy odebranej dla kąta ok. 70° w płaszczyźnie azymutu dla każdego kąta elewacji. Wynika to z tego, że pomiary były wykonywane w nieidealnym środowisku na wyniki te mogą składać się również komponenty wielodrogowości - odbicia od ścian itp. Na wykresach otrzymanych dla wzorca 1 oraz 2 można zauważyć, że rozkład mocy odbieranej jest podobny, przy czym dla wzorca drugiego, czyli dla wszystkich elementów RMA włączonych, odbierany sygnał jest mocniejszy. W przypadku wzorca 1 dla kąta -9° elewacji występują 2 minima charakterystyki, natomiast dla drugiego wzorca pojawia się po jednym minimum dla kątów 9° i -9°. Dla wzorca 3 możemy zaobserwować większą liczbę minimów dla kąta elewacji równego -9°. Ciekawym przypadkiem wzmocnienia sygnału cechuje się wzorzec 4 dla kąta elewacji -18° i 90° kąta azymutu.

W kierunku tym występuje spore wzmocnienie sygnału względem centralnego punktu wykresu. Dla kątów elewacji -9° i 9° zauważamy takie kąty azymutu, dla którch moc sygnału odbieranego osłabiona jest o ponad 20 dB względem maksymalnego położenia. Losowo wygenerowany wzorzec stworzył sytuację, w której na kącie elewacji -27° pojawiły się dwa wyraźne osłabienia sygnału na kącie 86° i 111° o ok. o 10dB. 
\section{Zakończenie}
W pracy przedstawiono wyniki pomiarów dwu- i trójwymiarowej charakterystyki odbiciowej modułów RMA w warunkach laboratoryjnych w środowisku typowym. Uzyskane wyniki pokazują dużą zmienność charakterystyk w zależności od wartosci kąta azymutu oraz elewacji. Wahania mocy mogą dochodzić nawet do 20 dB. Pozwala to stwierdzić, że istnieje możliwość wykorzystania wzorców RMA jako swoistej książki kodowej. 
\section{Podziękowania}
Praca powstała w ramach projektu badawczego OPUS 2021/43/B/ST7/01365 finansowanego przez Narodowe Centrum Nauki w Polsce.

\end{multicols}

\end{document}